\documentclass[preprint,doublespacing,tightenlines,showpacs,showkeys,amsmath,endfloats*,amssymb,a4paper,12pt]{revtex4}
\usepackage{epsfig}
\usepackage{amssymb}
\usepackage{amssymb}
\usepackage{amsmath}
\usepackage{color}
\usepackage{subfigure}
\usepackage{psfrag}

\preprint{Physics of Fluids}

\begin{document}
\title{A volume-based hydrodynamic approach to sound wave propagation in a monatomic gas} %
\author{S.\ Kokou Dadzie}
\email{ kokou.dadzie@strath.ac.uk}
\author{Jason M.\ Reese}
\email{jason.reese@strath.ac.uk}
\affiliation{Department of Mechanical Engineering, University of Strathclyde,\\
Glasgow G1 1XJ, UK}

\date{\today}

\begin{abstract}
We investigate sound wave propagation in a monatomic gas using a
volume-based hydrodynamic model. In reference
\cite{DadzieReese.PhysicaA.2008}, a microscopic volume-based kinetic
approach was proposed by analyzing molecular spatial distributions;
this led to a set of hydrodynamic equations incorporating a
mass-density diffusion component. Here we find that these new
mass-density diffusive flux and volume terms mean that our
hydrodynamic model, uniquely, reproduces sound wave phase speed and
damping measurements with excellent agreement over the full range of
Knudsen number. In the high Knudsen number (high frequency) regime,
our volume-based model predictions agree with the plane standing
waves observed in the experiments, which existing kinetic and
continuum models have great difficulty in capturing. In that regime,
our results indicate that the ``sound waves'' presumed in the
experiments may be better thought of as ``mass-density waves'',
rather than the pressure waves of the continuum regime.
\end{abstract}


\maketitle
\newpage

\section{Introduction}
One of the assumptions underpinning the conventional
Navier-Stokes-Fourier set of equations is that of local
thermodynamic equilibrium. This assumption allows the representation
of thermodynamic variables (e.g.\ temperature, density, pressure) as
locally constant at a given time and position, and the use of
equations of state. The assumption that microscopic relaxation
processes are not of concern is, however, inadequate in flows where
the microscopic relaxation time is comparable to the characteristic
time of evolution of the macroscopic field variables. In the kinetic
theory of dilute gases, such flows are identified with high Knudsen
numbers (conventionally defined as a ratio of the average time
between molecule/molecule collisions to a macroscopic characteristic
time of the flow, however see
\cite{LockerbyReeseStruchtrup.ProcRoySoc.2009}). Experimental
observations of sound wave propagation at high Knudsen number
challenge many continuum hydrodynamics and kinetic theory models
\cite{cercignani_rouge,cercignani-complet,Greenspan_1950,Meyer_Sessler1957};
it is well-known that the Navier-Stokes-Fourier model fails to
predict sound wave propagation at high Knudsen number. Another
problem arises in the so-called ``heat conduction paradox'',
according to which an unphysical infinite speed of thermal wave
propagation is predicted by the energy equation closed with
Fourier's law.

Generally, techniques for investigating gas flows in which the
Navier-Stokes-Fourier model is inadequate are based on approximate
solutions to the Boltzmann dilute gas kinetic equation, for which a
wide number of mathematical methods are found in the literature
\cite{cercignani-complet}. Regarding the specific problem of
predicting sound wave propagation in monatomic gases in the high
Knudsen number regime, many of these Boltzmann based approximations
fail, as does Navier-Stokes-Fourier
\cite{cercignani-complet,Greenspan_1950,Meyer_Sessler1957,Kanhn_Mintzer_1965_PhyofFlu,StruchtrupTorrilhon2003PhyFlu}.
While a few have shown some agreement with experiments
\cite{schotter:1974,Buckner_PhyFlu1966}, detailed analysis makes any
conclusion far from clear-cut
\cite{cercignani-complet,Paul-Dellar.PhyofFlu2007,Sirovich_thurber1965,Loyalka_Cheng_PhysFlu_1979}.
For example, if the experimental set-up is configured to measure
propagations of plane harmonic waves \cite{schotter:1974}, Boltzmann
kinetic models predict unconventional pressure fields, even though
the phase speeds and damping coefficients do agree with the
experimental data \cite{Buckner_PhyFlu1966}. Recently developed
continuum models also show discrepancies in these predictions,
particularly in the damping
\cite{Spiegel_Thiffeault2003,Paul-Dellar.PhyofFlu2007}.

The unphysical predictions of the conventional Navier-Stokes-Fourier
model have been investigated in terms of the ``heat conduction
paradox''. Early investigations criticized the expression of
Fourier's law, suggesting instead that the heat flux expression
should be transformed from the parabolic form of the heat conduction
equation to a simple hyperbolic equation with a finite speed of
propagation. While the original demonstration by Cattaneo
\cite{Cattaneo_CRAS_1958} has a flaw \cite{Ingo_muller_2001}, a
Cattaneo-Vermot heat flux has been formalized more elegantly using
fading memory theory (which essentially aims to remove the local
equilibrium assumption). Variants and generalizations have been
proposed, and compatibility with the second law of thermodynamics
has been assessed \cite{Petrov_Szekeres_2008,zhuomin2007}. However,
these investigations concentrate on modifications to the simple heat
conduction equation; they are not, to our knowledge, developed
within the framework of complete fluid dynamic equations and a full
dispersion analysis.

In this paper we investigate hydrodynamic models in which the
assumptions limiting the application of the conventional
Navier-Stokes-Fourier model are clearly released; this is therefore
outside the framework of pure approximation solutions to the
Boltzmann kinetic equation. In previous work, we proposed releasing
the local equilibrium assumption by including the spatial
distributions of molecules within the kinetic description
\cite{DadzieReese.PhysicaA.2008}. While our description was
motivated by an unusual volume diffusion claimed by Brenner
\cite{Brenner-Phys.vol2005,Brenner.PhysicaA.revs.2005}, it has been
recently pointed out that the original Brenner modification does not
predict sound wave speeds correctly
\cite{Marques_ChinePhyLet_2008,Greensield_reese_2007}.

Here we show that our volume-based hydrodynamic model can reproduce
the experimental sound wave propagation data from
ref.~\cite{Meyer_Sessler1957} with excellent agreement. Moreover,
our model offers a more reliable explanation of the experiments,
which were designed to range up to the free molecular regime in
which there are no collisions between molecules and therefore the
definition of sound as a pressure wave becomes problematic.

This paper starts with a summary of our volume model that
incorporates effects from microscopic spatial distributions of the
gaseous molecules. Subsequently, a linear stability analysis of the
model equations is performed, and the predicted dispersion and
damping compared with experiments.

\section{Summary of the volume-based hydrodynamic description}
The traditional single particle distribution function used in the
Boltzmann kinetic equation for a monatomic gas attributes no
particular importance to the spatial arrangements of molecules. An
average number of molecules is associated with a position $X$ and a
velocity $\xi$. In order to account for microscopic spatial
fluctuations, due to non-uniformity in molecular spatial
configurations, we have considered within the set of microscopic
variables the microscopic free volume, $v$, around each gaseous
molecule. A single particle distribution function $f(t, X,\xi, v )$
is then defined to describe the probability that a molecule at a
given time $t$ is located in the vicinity of position $X$, has its
velocity in the vicinity of $\xi$, and has around it a microscopic
free space given by the additional variable $v$.

A Boltzmann-like kinetic equation for $f(t, X, \xi, v)$ is then
derived as \cite{DadzieReese.PhysicaA.2008}:
\begin{equation}
\label{eq.boltzmann.kok}
\frac{\partial f}{\partial t} + (\xi \cdot \nabla ) f  + W  \frac{\partial f}{\partial v} = \int \int (f^+
f_1^+ - f f_1) \sigma \xi_r d_{\omega} d_{\xi_1} ,
\end{equation}
in which the term on the right-hand-side is the hard sphere molecule
collision integral; $f = f(t, X, \xi, v )$ and $f_1 = f(t, X, \xi_1,
v_1 )$ refer to post-collision molecules, $f^+ = f(t, X, \xi^+, v^+
)$ and $f_1^+ = f(t, X, \xi_1^+, v_1^+ )$ refer to pre-collision
molecules, $\xi_r = \xi-\xi_1 $ is the molecule relative velocity,
$\sigma$ the collision differential cross section, $d_\omega$ an
element of solid angle. On the left-hand-side appears a new term
involving $W$, which arises primarily from the introduction of the
new variable $v$ into the distribution function. In the derivation
of equation (\ref{eq.boltzmann.kok}), molecular exchanges of
momentum through interactions have been assumed to be independent of
their spatial configurations.

Three contributions to the time variations of  $f(t, X, \xi, v )$
are seen within equation (\ref{eq.boltzmann.kok}). Molecular
free-stream motions are given by the second term on the
left-hand-side. The third term on the left-hand-side arises from
effects of molecular interactions on their spatial distributions.
Finally, the collision integral is the traditional momentum exchange
between molecules that provides changes in molecular velocities.
These latter two terms infer that the real molecular potential
interactions are represented in this kinetic model by two separate
actions: intermolecular force effects on spatial distributions, and
collisional effects on molecular velocities.

\subsection{Molecular average properties}
As $f(t,X,\xi,v)$ is defined as a probability density function,
we have a normalization factor,
\begin{equation}
A_n(t,X) =  \int_{-\infty}^{+\infty}  \int_0^{+\infty} f(t, X, \xi, v
)  d_v d_\xi \ .
\end{equation}
The mean value, $\bar{Q}(t,X)$, of a gas property $Q$ is then
defined by,
\begin{equation}
\label{meanvalu}
\bar{Q}(t,X)   = \frac{1}{A_n(t,X)}  \int_{-\infty}^{+\infty}  \int_0^{+\infty} Q  f(t, X, \xi, v )
d_v d_\xi \ .
\end{equation}
The local average of $v$ is therefore the local mean free volume
$\bar{v}(t,X)$ around each gaseous molecule, i.e.
\begin{equation}
\label{meanvoldefi}  \bar{v} (t,X) = \frac{1}{A_n(t,X)}\int_{-\infty}^{+\infty}
\int_0^{+\infty} v f(t, X, \xi, v )  d_v d_\xi \  .
\end{equation}
From this mean value of the volume around a molecule, we define the
mass-density in the vicinity of position $X$ through:
\begin{equation}
\label{massdensitydef}
\bar{\rho} (t,X)  =
\frac{M}{\bar{v}(t,X)}  \ ,
\end{equation}
where $M$ is the molecular mass. Two mean velocities are defined
using two different weighting values: the local mean mass-velocity,
$U_m (t,X)$, is given through
\begin{equation}
\label{vitessemass}
A_n(t,X) U_m (t,X) = \int \int \xi f(t, X, \xi, v ) d_\xi d_v ,
\end{equation}
and a local mean volume-velocity, $U_v(t,X)$, by using the
microscopic free volume as the weighting,
\begin{equation}
\label{vitessevolume}
\bar{v} (t,X) A_n(t,X)  U_v (t,X) = \int \int  v \xi f(t, X, \xi, v ) d_\xi d_v .
\end{equation}
The  two definitions $U_v$ and $U_m$ coincide if $v$ is a constant,
i.e. in a homogeneous medium where density is constant throughout.
It can be shown that the difference between these two velocities,
$U_v - U_m = \bar{v}^{-1}\mathbf{J}_v$, behaves like a mass-density
diffusion \cite{DadzieReese.PhysicaA.2008}.

\subsection{A volume-based hydrodynamic set of equations}
Hydrodynamic equations are derived as conservation equations
obtained from the kinetic equation, accounting for a
reclassification of convective/diffusive fluxes required by the
appearance of the two different velocities. The set of equations is
obtained \cite{DadzieReese.PhysicaA.2008}:
\begin{description}
\item[Continuity]
\begin{eqnarray}
\label{massnewhydro} \frac{D A_n }{ Dt} = -  A_n \nabla \cdot U_m \ ,
\end{eqnarray}
\item[Mass-density]
\begin{eqnarray}
\label{densitynewhydro}
A_n  \frac{D \bar{v} }{ D t}
=- \nabla \cdot
[A_n \mathbf{J}_v] +  A_nW  ,
\end{eqnarray}
\item[Momentum]
\begin{eqnarray}
\label{momentumnewhydro} A_n \frac{D U_m }{ D t}  = -  \nabla \cdot A_n
\left(\mathbf{P'} -  \frac{1}{\bar{v}^2}\mathbf{J}_v \mathbf{J}_v
\right) ,
\end{eqnarray}
\item[Energy]
\begin{align}
\label{energyhydro}
 A_n \frac{D }{ D t} \left[ \frac{1}{2} U_m^2 + e'_{in} - \frac{1}{2\bar{v}^2}\mathbf{J}_v^2 \right] & = - \nabla
\cdot A_n \left[ \left(\mathbf{P'} - \frac{1}{\bar{v}^2}\mathbf{J}_v
\mathbf{J}_v \right)\cdot U_m\right]
\\ \nonumber
 &   - \nabla \cdot A_n \left[\mathbf{q'}
+\frac{1}{\bar{v}}\mathbf{P'} \cdot \mathbf{J}_v
   + \frac{1}{\bar{v}}\left( e'_{in} -  \frac{1}{\bar{v}^2}\mathbf{J}_v^2\right
   )\mathbf{J}_v\ \right] \ .
\end{align}
\end{description}
where we denote the material derivative $D/Dt \equiv
\partial /\partial t + U_m \cdot~\nabla$. The flow variables are:
the probability density $A_n$ (which is, however, not a physical
property), the mass-density $\bar{\rho}$, the mass-velocity $U_m$,
and the internal energy $e'_{in}$.

Following, provisionally, the classical phenomenological Fick's law
for a diffusive flux, the model may be closed by the constitutive
relations:
\begin{align}
\label{fluxes-new} \frac{M \mathbf{P'}_{ij}}{\bar{v}} & =  p' \delta_{ij} -  \mu'
\left( \frac{\partial U_{v_i}}{\partial X_j} + \frac{\partial
U_{v_j}}{\partial X_i}\right) + \eta' \frac{\partial
U_{v_k}}{\partial X_k}\delta_{ij} \ ,
 \\
\frac{M\mathbf{q'}}{\bar{v}}  & =  - \kappa'_h  \nabla T' \ ,  \\
 \mathbf{J}_v & = - \kappa_m    \nabla \bar{v}  \ ,
 \label{fluxes-new-end}
\end{align}
in which we have defined $M e'_{in}= (3/2) kT'$  with $T'$ being the
kinetic temperature, or $p'= (2/3) \bar{\rho} e'_{in} $ with $p'$
being the kinetic pressure, and $U_v = U_m +
\bar{v}^{-1}\mathbf{J}_v$. The coefficients $\mu'$, $\kappa'_h$,
$\eta'$ and  $\kappa_m$ are, respectively, dynamic viscosity, heat
conductivity, bulk viscosity, and the mass-density diffusion
coefficient. As the kinetic pressure $p'$ is defined by the trace of
the pressure tensor we also have $\frac{2}{3}\mu' - \eta'=0 $.

Previous volume diffusion hydrodynamic models have been based on
separating the mean velocity in the conventional mass conservation
equation (continuity equation), from the mean velocity in the
Navier-Stokes momentum equation via Newton's viscosity law
\cite{Brenner.PhysicaA.revs.2005}. This has proven controversial
\cite{Mario_PRL} --- problems in differentiating the mass-flux from
the momentum density, and in conserving angular momentum when the
velocity on the left-hand-side of the Navier-Stokes equation is
substituted for, have been raised. In our approach, however, a mass
flux is given by $\bar{\rho}U_v$ from the mass-density equation
(\ref{densitynewhydro}), and involves the same velocity, $U_v = U_m
+ \bar{v}^{-1}\mathbf{J}_v$, as in Newton's viscosity law (equation
\ref{fluxes-new}). Meanwhile, the velocity on the left-hand-side of
the new momentum equation (\ref{momentumnewhydro}) remains the
conventional mass velocity $U_m$ (following Newton's second law).
Consequently the two flaws mentioned in connection with volume-based
hydrodynamics in reference \cite{Mario_PRL} are not present in our
set of equations (\ref{massnewhydro})--(\ref{fluxes-new-end}).

\subsection{The localized rate of change of volume, $W$}
A consequence of our localized microscopic volume description is the
appearance of $W$, the time rate of change of microscopic volume.
Although this term could be proposed using details of the
interactions between particles, here we instead test a
phenomenological expansion of $W=\delta v/ \delta t$ as a function
of the fluid macroscopic thermodynamic variables. First we relate
variations of the microscopic $v$ to variation of its macroscopic
average $\bar{v}$, through a relaxation approximation:
\begin{equation}\label{approx-property-fun}
\frac{\delta v}{\delta t} = \frac{d}{ dt} \left(\bar{v}+ \tau_s \frac{d \bar{v}}{d t}\right)\ .
\end{equation}
The derivative $\delta / \delta t$ refers to the time rate of change
of microscopic properties while $d/dt$ refers to the time rate of
change of macroscopic properties, with $\tau_s$ a relaxation time.
Expanding $d\bar{v}$ as a function of thermodynamic variables we
have:
\begin{equation}\label{expres_W}
 \frac{1}{\bar{v}} W= \alpha \frac{d T'}{d t} + \beta \tau_s \frac{d^2 T'}{d t^2}   - \chi \frac{d p'}{dt} - \gamma \tau_s \frac{d^2 p'}{dt^2},
\end{equation}
where $\alpha$, $\beta$, $\chi$, $\gamma$ are the gas expansion and
compressibility coefficients given by,
\begin{equation}\label{expres-alpha-chi}
\alpha = \left(\frac{1}{\bar{v}}\frac{\partial \bar{v}}{\partial
T^{\prime}}\right)_{p^{\prime}}, \hspace{2.5em}
\chi = -
\left(\frac{1}{\bar{v}}\frac{\partial \bar{v}}{\partial
p^{\prime}}\right)_{T^{\prime}} \ ,
\end{equation}
and
\begin{equation}\label{expres-beta-gamma}
\beta = \left(\frac{1}{\bar{v}}\frac{\partial^2 \bar{v}}{\partial
{T^{\prime}}^2}\right)_{p^{\prime}}, \hspace{2.5em}
\gamma = -
\left(\frac{1}{\bar{v}}\frac{\partial^2 \bar{v}}{\partial
{p^{\prime}}^2}\right)_{T^{\prime}} \ .
\end{equation}
In our description local thermodynamic equilibrium is not required.
Relations $M e'_{in}= (3/2) kT'$ and $p'= (2/3) \bar{\rho} e'_{in} $
define the temperature and pressure (following their classical
definitions in kinetic theory), therefore there is a reciprocal
relation between temperature and pressure, $p'= kT'/\bar{v}$, by
construction without further assumption. If the perfect gas
(equilibrium) equation of state is enforced, and we confuse $\delta
v / \delta t$ with $d\bar{v}/dt$ in equation (\ref{expres_W}), then
the gas expansion and compressibility coefficients in equations
(\ref{expres-alpha-chi}) are the ideal gas coefficients, i.e.
$\alpha=1/T'$ and  $\chi =1/p'$, and the second order contributions
vanish from equation (\ref{expres_W}). But as we are not restricting
ourselves to local thermodynamic equilibrium, a departure from these
ideal coefficients may be expected.

Now we turn to investigate sound dispersion using both the first and
the second order approximations to $W$ given in equation
(\ref{expres_W}).

\section{Linear stability analysis and sound wave propagation}
\subsection{Linearized one-dimensional equations \label{sec-Linearequation}}
We consider our hydrodynamic model in a one-dimensional flow
configuration. An equilibrium ground state is defined by the flow
variables $ A_n^0$, $\bar{\rho}^0$,  $T^0$, $p^0=R\bar{\rho}^0T^0$,
U$_m^0= U_v^0=0$, with $R$ the specific gas constant. Then a
perturbation from this ground state is introduced as follows:
\begin{align}\label{charateristics}
& A_n = A_n^0(1+A_n^*), \quad \bar{\rho} = \bar{\rho}^0(1+\rho^*),
\quad T' =
T^0(1+T^*), \\
& U_m = U_m^* \sqrt{RT^0}, \quad p' = p^0(1+p^*), \notag
\end{align}
where the asterisked variables represent dimensionless quantities.
The perturbation of the volume velocity is specified through the
relationship $U_v = U_m + \bar{v}^{-1}\mathbf{J}_v$. Linearizing
$p'= kT'/\bar{v}$ gives $p^* = \rho^* + T^* $. The dimensionless
space and time variables are given by,
 \begin{align}\label{micro-timedefo}
x = L x^*,  \ \  t =\frac{L}{\sqrt{RT^0}} t^* =\tau t^*,
\end{align}
with $\tau=L/\sqrt{RT^0}$. The dimensionless linearized equations,
including the general expression for $W$ in equation
(\ref{expres_W}), can therefore be written:
\begin{description}
\item[Continuity]
\begin{align}
\label{massnewhydrohat}
\frac{ \partial A_n^* }{ \partial t^*} +\frac{ \partial U_m^* }{ \partial x^*} =0 \ ,
\end{align}
\item[Mass-density]
\begin{align}
\label{density-trans12hat}
 \left(1- \chi^*\right)\frac{ \partial \rho^*}{\partial  t^*} - \kappa_m^* \frac{ \partial^2 \rho^*}{\partial {x^*}^2}+
 \left(\alpha^*-\chi^*\right) \frac{ \partial T^*}{\partial t^*}
 - \gamma^* \frac{ \partial^2 \rho^*}{\partial {t^*}^2}
 + \left(\beta^*-\gamma^*\right) \frac{ \partial^2 T^*}{\partial {t^*}^2}
  =0 ,
\end{align}
\item[Momentum]
\begin{align}
\label{momentumnewhydrohat}
 \frac{ \partial U_m^*}{\partial t^*}  - \frac{4}{3}\mu^* \frac{ \partial^2 U_m^*}{\partial {x^*}^2}
+
 \frac{ \partial A_n^* }{ \partial x^*} +  \frac{\partial T^*}{\partial x^*}  - \frac{4}{3}\mu^*\kappa_m^* \frac{ \partial^3 \rho^*}{\partial {x^*}^3} =0 ,
\end{align}
\item[Energy]
\begin{align}
\label{energyhydrohat}
\frac{ \partial T^*}{\partial t^*} +\frac{2}{3} \frac{ \partial U_m^*}{\partial x^*} -
\frac{2}{3}\kappa_h^* \frac{ \partial^2 T^*}{\partial {x^*}^2}+
\frac{5}{3}\kappa_m^* \frac{ \partial^2 \rho^*}{\partial {x^*}^2} =0 \ ,
 \end{align}
\end{description}
where the different dimensionless transport coefficients are given through:
 \begin{align}
 \label{transpo-coeff-dimens}
  \mu' = \bar{\rho^0}L \sqrt{RT^0}\mu^*,    \ \    \kappa_m =L \sqrt{RT^0} \kappa_m^*, \ \  \kappa_h' =
 \frac{L\bar{\rho^0}    (\sqrt{RT^0})^3} { T^0} \kappa_h^* ,
\end{align}
and
\begin{align}
 \label{transpo-coeff-dimensxiphi}
   \alpha = \frac{1}{T^0}\alpha^* ,  \  \ \chi = \frac{1}{p^0}\chi^* , \  \ \beta = \frac{1}{T^0}\beta^* ,  \  \  \gamma = \frac{1}{p^0}\chi^* .
\end{align}
Note that the dimensionless transport coefficients in equations
(\ref{transpo-coeff-dimens}) follow from the dimensionless form of
the hydrodynamic set of equations. Instead of using these
dimensionless coefficients, however, it may be more convenient to
use conventional parameters, i.e.\ the Knudsen number $K_n$, the
Prandtl number $P_r$, and an additional parameter $S_c$ that
involves the mass-density diffusivity. These are given by (denoting
$\mu^0 = \bar{\rho^0}L \sqrt{RT^0}$):
\begin{align}
 \label{dimensionless-number}
  K_n=  \frac{\mu' \sqrt{RT^0}}{p^0L} \equiv  \mu^*,   \ \    \frac{1}{S_c}=\frac{\kappa_m \bar{\rho^0}}{\mu^0} \equiv \kappa_m^*, \ \  \frac{1}{P_r}  = \frac{2}{5}
 \frac{\kappa_h'} {R\mu^0}  \equiv\frac{2}{5} \kappa_h^*.
\end{align}

We assume the disturbances $A_n^*$, $\rho^*$, $T^*$  and $U_m^*$ to
be wave functions of the form:
\begin{equation}
\phi^* =\phi_a^*  \exp\left[i\left( \omega t^* - K x^* \right) \right],
\label{harmonic-wrong}
\end{equation}
where $\omega$ is the complex wave frequency, $K$ is the complex
wave number, and  $\phi_a^*$ is the complex amplitude, so that:
\begin{equation*}
\frac{\partial \phi^*}{\partial t^*}  = i \omega \phi^*, \quad
\frac{\partial^2 \phi^*}{\partial {t^*}^2} = - \omega^2 \phi^*,
\quad \frac{\partial \phi^*}{\partial x^*} = -i K \phi^*, \quad
\frac{\partial^2 \phi^*}{\partial {x^*}^2} =   - K^2 \phi^*, \quad
\frac{\partial^3 \phi^*}{\partial {x^*}^3}  = i K^3 \phi^*.
\end{equation*}
The linearized hydrodynamic set of equations then yields the homogeneous system,
\begin{align}
\Xi (\omega,K)\times
\begin{Bmatrix}
A_n^*  \\
\rho^*   \\
  T^* \\
U_m^*
\end{Bmatrix}
= 0 ,
\end{align}
where
\begin{align}
\Xi(\omega,K)=
\begin{Bmatrix}
 i \omega     & 0                                                       & 0                                                           & -i K \\
 0       & \kappa _m^* K^2+i  \omega (1-\chi^* ) -\gamma^*\omega^2  \  &  \  i \omega (\alpha^* -\chi^* )+ (\beta^* -\gamma^* )\omega^2  & 0 \\
 0       & -\frac{5}{3} K^2 \kappa _m^*  & \frac{2}{3} \kappa _h^* K^2+i \omega     & -\frac{2 }{3}i K \\
 - i K     & -\frac{4}{3} i K^3 \mu^* \kappa _m^*        & -i K                                    & \frac{4 }{3} \mu^* K^2 +i \omega
\end{Bmatrix}.
\end{align}
The corresponding dispersion relation, obtained when the determinant
of $\Xi(\omega,K)$ is zero, is
\begin{align}
\label{dispersion} \nonumber \left[\frac{20 i \omega K_n K^4}{9
P_r}+\frac{5 K^4}{3 P_r}+\frac{5}{3} i \omega K^2-\frac{4}{3}
\omega^2 K_n K^2-\frac{5 \omega^2 K^2}{3 P_r}-i w^3\right]\times & \\
\nonumber
   \left[-\gamma^* \omega^2+i \left(1-\chi^*\right) \omega+\frac{K^2}{S_c}\right]-\left[\left(\beta^*-\gamma^*\right) \omega^2+i \left(\alpha^*
   -\chi^*\right) w\right] \times & \\
   \left[-\frac{4 i \omega K_n K^4}{3 S_c}-\frac{5 K^4}{3 S_c}+\frac{5 \omega^2 K^2}{3 S_c}\right] &=0 .
  \end{align}

\subsection{Dispersion and damping predictions compared with experiment}
When analyzing the dispersion and stability characteristics of our
model, we compare our results for sound propagation in argon gas
with experimental data from reference \cite{Meyer_Sessler1957}.

Choosing the harmonic wave expression (\ref{harmonic-wrong}) is in
line with previous analysis of this problem, and the dimensionless
phase speed $\Upsilon_l$, and dimensionless spatial damping
$\Lambda_l$, are then commonly defined by
\cite{Greenspan_1950,Meyer_Sessler1957,Paul-Dellar.PhyofFlu2007}:
\begin{align}\label{damp-speed-expre}
  \frac{1}{\Upsilon_l}  =  \sqrt{\frac{5}{3}}\frac{Re[K]}{\omega} ,  \ \  \Lambda_l = -  \sqrt{\frac{5}{3}}\frac{Im[K]}{\omega} \ .
\end{align}
Setting the Knudsen number $K_n$, defined in equation
(\ref{dimensionless-number}), to $1$ makes our analysis agree with
that of Greenspan \cite{Greenspan_1950}, in which variations of
frequency $\omega$ are interpreted as variations of Knudsen number
(the limitations of this particular interpretation are outlined in
the Appendix to this present paper). Although more recent
experimental data with a different analysis exists, we choose this
approach first in order to make comparisons with previously
published works
\cite{Greenspan_1950,Meyer_Sessler1957,Paul-Dellar.PhyofFlu2007,StruchtrupTorrilhon2003PhyFlu}.

We also note here that a solution to a dispersion relation such as
equation (\ref{dispersion}) consists of various discontinuous
solutions generating a number of modes; one of these is expected to
correspond to the sound mode. In this paper, we include in our
results figures all modes, for the sake of a complete analysis.

Linear stability criteria are as follows
\cite{Greensield_reese_2007}: for the set of equations to be time
stable, $\omega(K)$ as a root of the dispersion relation
(\ref{dispersion}) should satisfy $Im[\omega(K)]\geq 0$ for all $K$
real. On the other hand, the set of equations will be stable in
space if $K(\omega)$ as a root of the dispersion relation satisfies
$Im[K(\omega)]\times Re[K(\omega)] < 0$ for all $\omega\geq0$.

\subsubsection{A first order approximation to $W$: $\beta^* = \gamma^*= 0$}
\begin{figure}[p]
     \centering
     \subfigure[\ Normalized inverse phase speed varying with $\omega^{-1}$]
     {\label{speed-navier}
                \includegraphics[width=0.8\textwidth,height=0.4\textwidth]{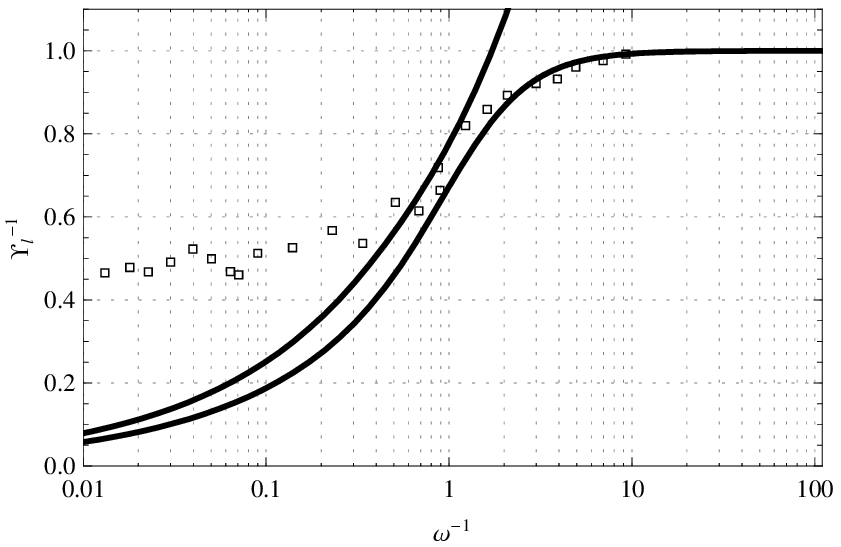}}
     \hspace{.3in}
     \subfigure[ \ Normalized damping coefficient varying with $\omega^{-1}$]
     {  \label{damp-navier}
         \includegraphics[width=0.8\textwidth,height=0.4\textwidth]{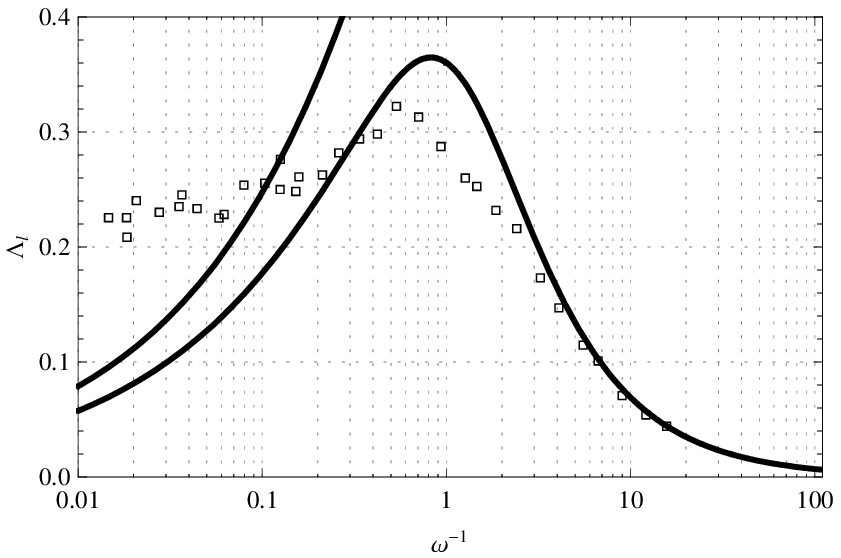}}\\
     \caption{
     Comparison of our volume-based dispersion predictions with experiments,
     with $W$ represented by a first order approximation, and using the definitions
     in equation (\ref{damp-speed-expre}). Experimental data are represented by the
     discrete squares. With $\alpha^*= \chi^*=1$ the dispersion relation is the
     same as for the Navier-Stokes-Fourier model.}
     \label{navier}
\end{figure}

First we set $\beta^* = \gamma^*= 0$, that is, $W$ is approximated
only by the first order terms in equation (\ref{expres_W}). For
$\alpha^* = \chi^*= 1$ the dispersion and stability characteristics
of our model correspond to those of the Navier-Stokes-Fourier model.
The equations are also stable in both time and space. Figure
\ref{navier} shows both the inverse phase speed and the damping as a
function of inverse frequency (i.e.\ inverse Knudsen number),
compared with experimental data \cite{Meyer_Sessler1957}.
Navier-Stokes-Fourier has only two modes: one mode fits the phase
speed and damping measurements at low Knudsen number, but has an
infinite speed of propagation for high Knudsen number. The second
mode shows an infinite inverse phase speed at low Knudsen number,
and is interpreted as the heat mode
\cite{Buckner_PhyFlu1966,Meyer_Sessler1957}.

\begin{figure}[p]
     \centering
     \subfigure[\ Temporal stability; $\alpha^*= 0.28$, $\chi^*=0.48$, $S_c=0.9$]
     {\label{stab-time-navier-corrected}
                \includegraphics[width=0.8\textwidth,height=0.4\textwidth]{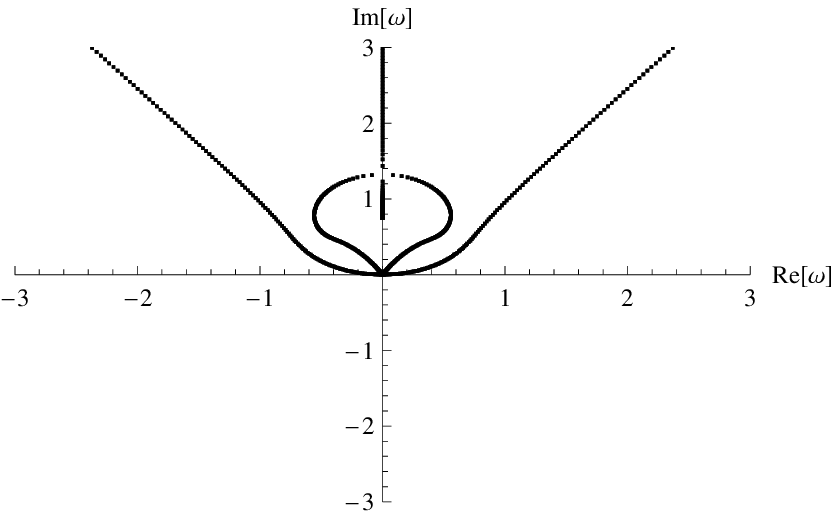}}
     \hspace{.3in}
     \subfigure[ \ Spatial stability; $\alpha^*=0.28$, $\chi^*=0.48$, $S_c=0.9$]
     { \label{stab-space-navier-corrected}
              \includegraphics[width=0.8\textwidth,height=0.4\textwidth]{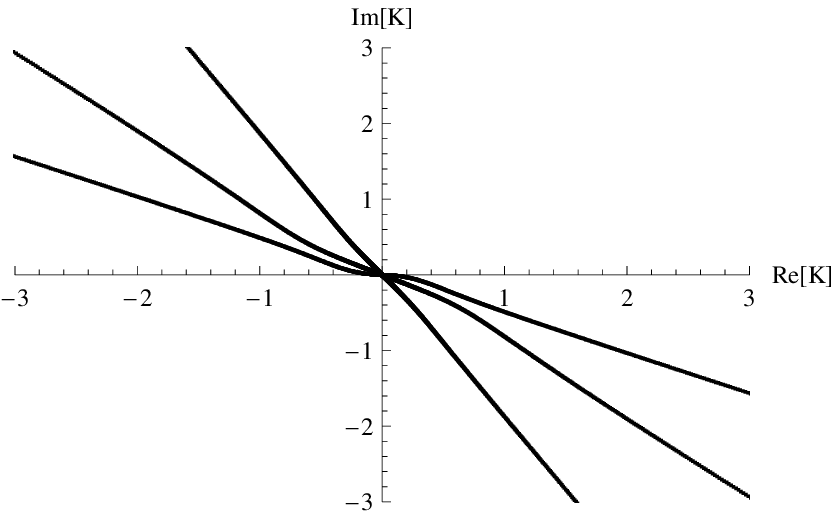}}\\
     \caption{Stability analysis of our volume-based hydrodynamic equations, with $W$ described by a first order approximation only.
      Our equations are stable in both space and time if ($\alpha^* \leq 1,\chi^*\leq 1$) or ($\alpha^* \geq 1 , \chi^* \leq 0.5$) }
     \label{stab-navier-corrected}
\end{figure}

\begin{figure}[p]
     \centering
     \subfigure[\ Normalized inverse phase speed varying with $\omega^{-1}$; $\alpha^*= 0.28$, $\chi^*=0.48$, $S_c=0.9$]
     {\label{speed-navier-corrected}
      \includegraphics[width=0.8\textwidth,height=0.4\textwidth]{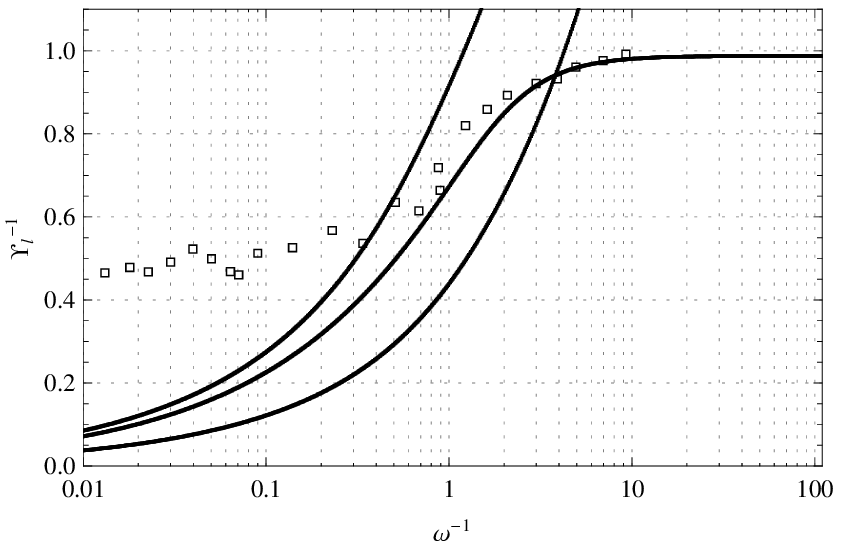}}
     \hspace{.3in}
     \subfigure[\ Normalized damping coefficient varying with $\omega^{-1}$; $\alpha^*=0.28$ and $\chi^*=0.48$, $S_c=0.9$]
     {\label{damp-navier-corrected}
              \includegraphics[width=0.8\textwidth,height=0.4\textwidth]{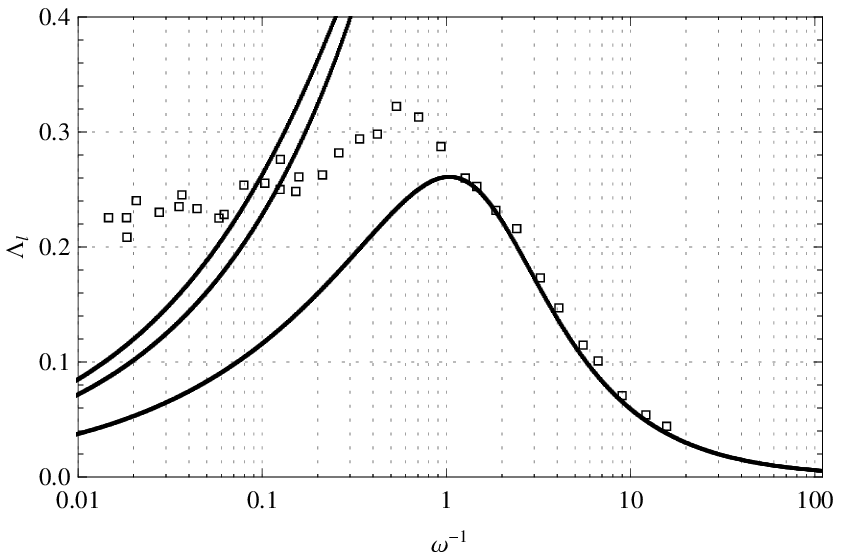}}\\
     \caption{
     Comparisons of our volume-based dispersion predictions with experiments,
     with $W$ represented by a first order approximation, and using the definitions
     in equation (\ref{damp-speed-expre}). Experimental data are represented by the
     discrete squares. Note the improvement on damping predictions compared to figure \ref{navier}.
     \label{navier-corrected}}
\end{figure}

Departures from these predictions are expected for our volume-based
hydrodynamic model when $\chi^* \neq \alpha^* $. We find that the
model is stable, in the case of a first order approximation to $W$,
if $\alpha^*$ and $\chi^*$ are both simultaneously smaller than one,
or $\alpha^* \geq 1$ and $\chi^*\leq0.5$, approximately; this is
illustrated in Figure \ref{stab-navier-corrected}. Comparison of the
dispersion with experiments shows globally the same results as in
the Navier-Stokes-Fourier case. But, as seen in figure
\ref{navier-corrected} where we have $\alpha^* =0.28$, $\chi^*
=0.48$ and $S_c=0.9$, the agreement with the low frequency regime is
improved, particularly in the damping coefficient. Both the phase
speed and the damping are adequately predicted up to $Kn=1$, whereas
the damping was predicted only up to $Kn=0.3$ by
Navier-Stokes-Fourier alone (figure \ref{damp-navier}).

Figure \ref{navier-corrected} also shows that there are now three
modes, two of which display transient diffusion behaviour (i.e.\
high damping in low frequency regimes). While one of these should be
considered as the heat mode, as previously, the other should be
attributed to transient mass-density diffusion, as introduced by our
new volume-based description (in addition to the heat diffusion).
This new mode is the most affected by the mass-density diffusivity,
i.e., by $S_c$. The high frequency regime is still incorrectly
predicted by the sound mode, as in the case of
Navier-Stokes-Fourier. Later we will see that the infinite speed of
propagation and zero damping in the high frequency regime can all be
removed with the inclusion of the new mass-density mode.

\subsubsection{A second order approximation to $W$, $\alpha^* = \chi^*= 0$}
\begin{figure}[p]
     \centering
     \subfigure[ \ Temporal stability; $\beta^*= 0.28$, $\gamma^*=0.48$, $S_c=0.14$]
     {\label{stab-kokou-time}
                \includegraphics[width=0.8\textwidth,height=0.4\textwidth]{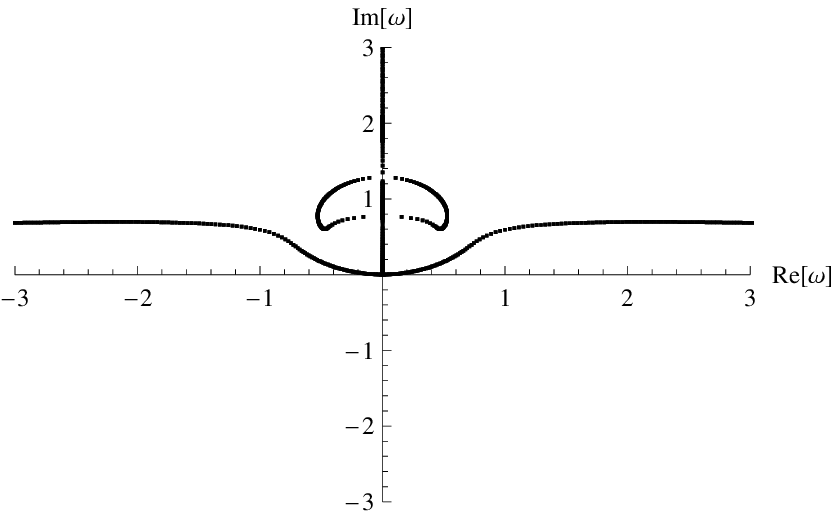}}
     \hspace{.3in}
     \subfigure[ \ Spatial stability; $\beta^*=0.28$, $\gamma^*=0.48$, $S_c=0.14$]
     { \label{stab-kokou-space}
              \includegraphics[width=0.8\textwidth,height=0.4\textwidth]{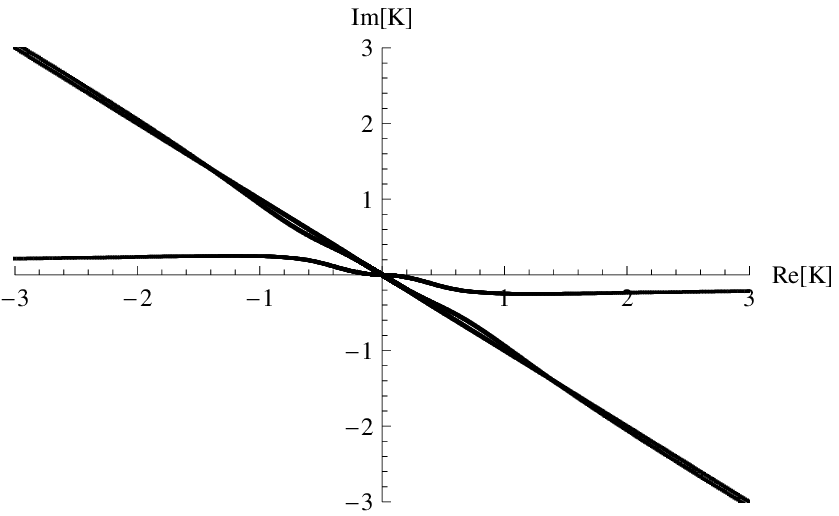}}\\
     \caption{Stability analysis of our volume-based hydrodynamic equations, with $W$ described by a second order approximation only.
Our equations are stable in both space and time if $0\leq \gamma^*
-\beta^*\leq 1.3$.}
     \label{stab-kokou-second}
\end{figure}

\begin{figure}[p]
     \centering
     \subfigure[ \ Normalized inverse phase speed varying with $\omega^{-1}$; $\beta^*=0.28$, $\gamma^*=0.48$, $S_c=0.14$]
     {\label{speed-kokou}
                \includegraphics[width=0.8\textwidth,height=0.4\textwidth]{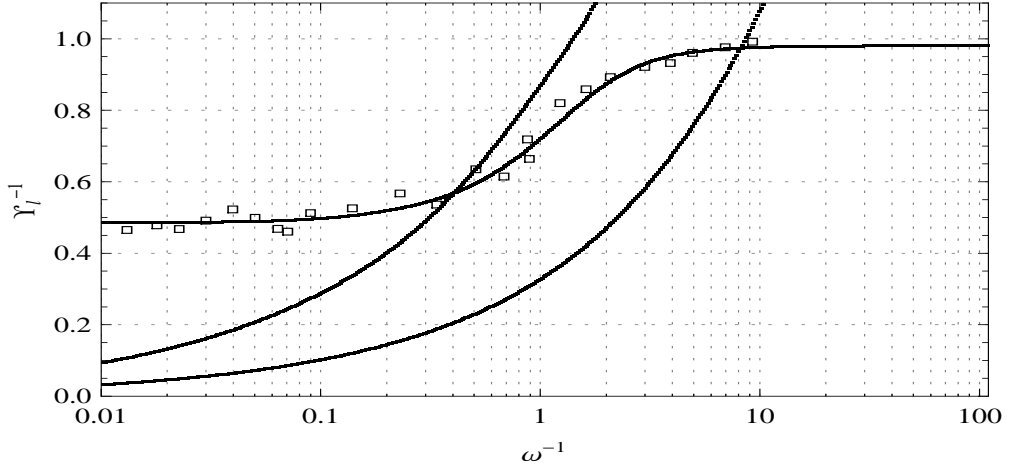}}
     \hspace{.3in}
     \subfigure[\ Normalized damping coefficient varying with $\omega^{-1}$; $\beta^*=0.28$, $\gamma^*=0.48$, $S_c=0.14$]
     {\label{damp-kokou}
     \includegraphics[width=0.8\textwidth,height=0.4\textwidth]{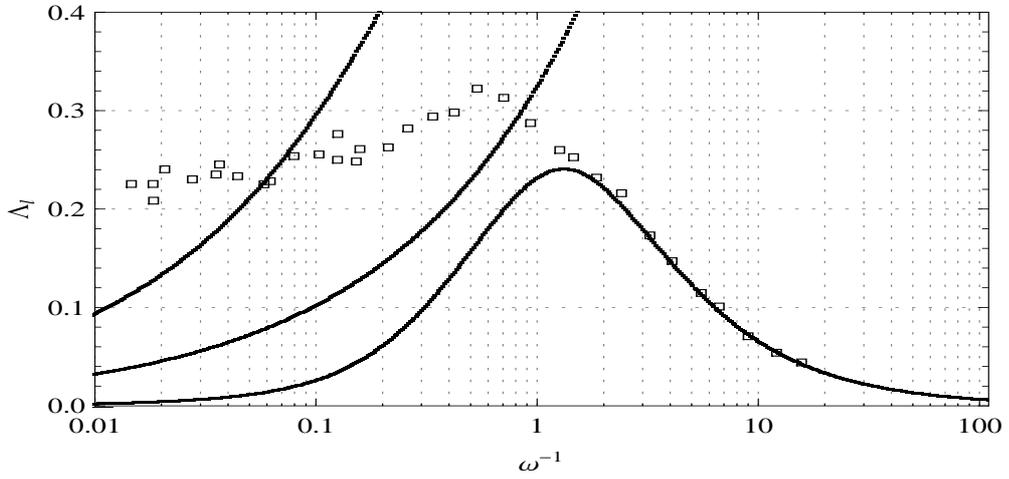}}\\
     \caption{Comparison of our volume-based dispersion predictions with experiments,
     with $W$ described by a second order approximation, and using equation (\ref{damp-speed-expre}).
     Experimental data are represented by the discrete squares. Note the agreement with the phase
     speed for all Knudsen numbers.}
     \label{kokou-second}
\end{figure}

\begin{figure}[p]
     \centering
     \vspace{-5in}
     \subfigure[ \ Inverse phase speed compared with other models]
     {\label{speedcomparedother}
                \includegraphics[width=0.78\textwidth,height=0.35\textwidth]{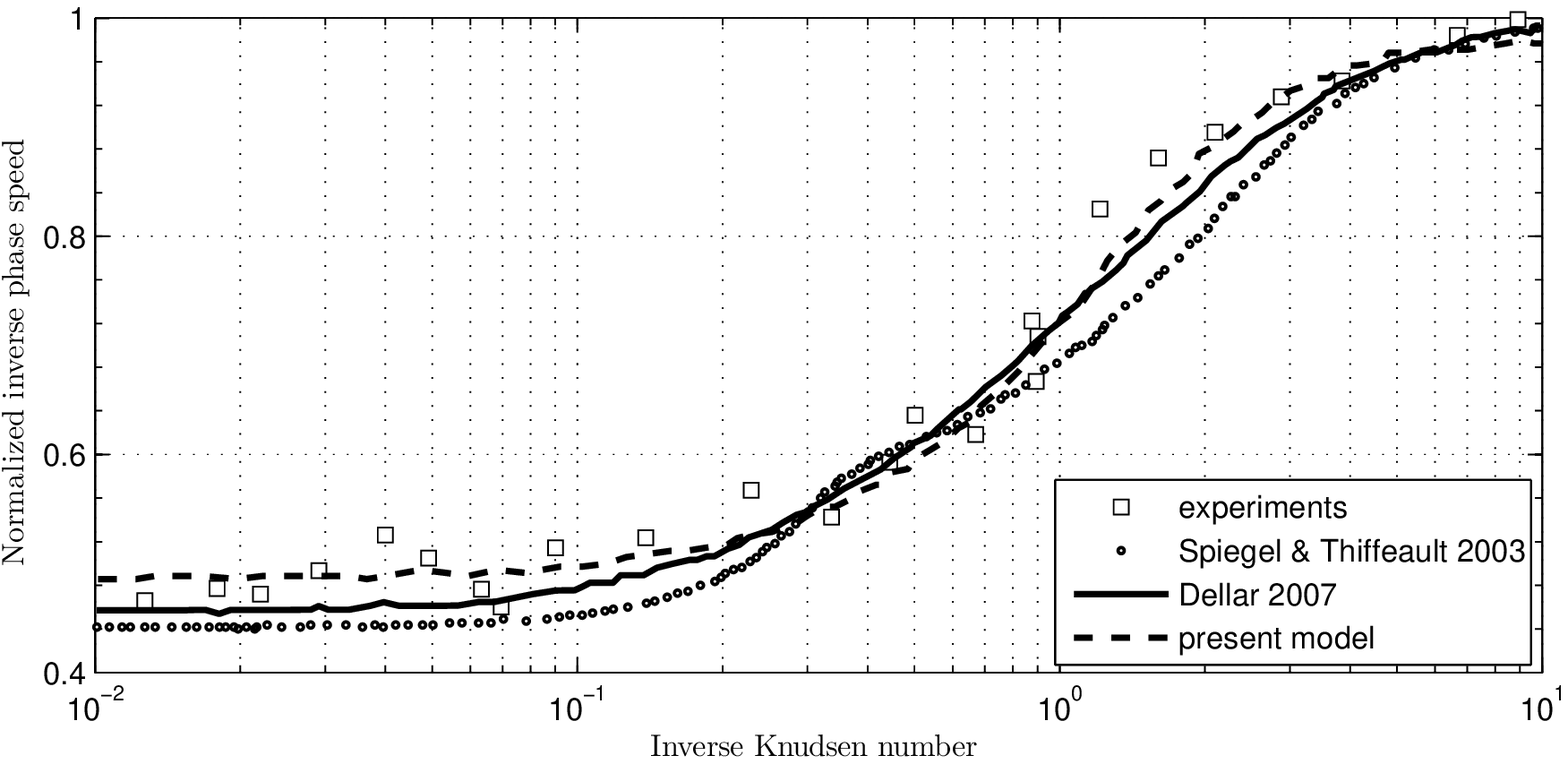}} %
   \vspace{1em}
     \subfigure[ \ Damping coefficient compared with other models]
     { \label{dampingcomparedother}
              \includegraphics[width=0.78\textwidth,height=0.35\textwidth]{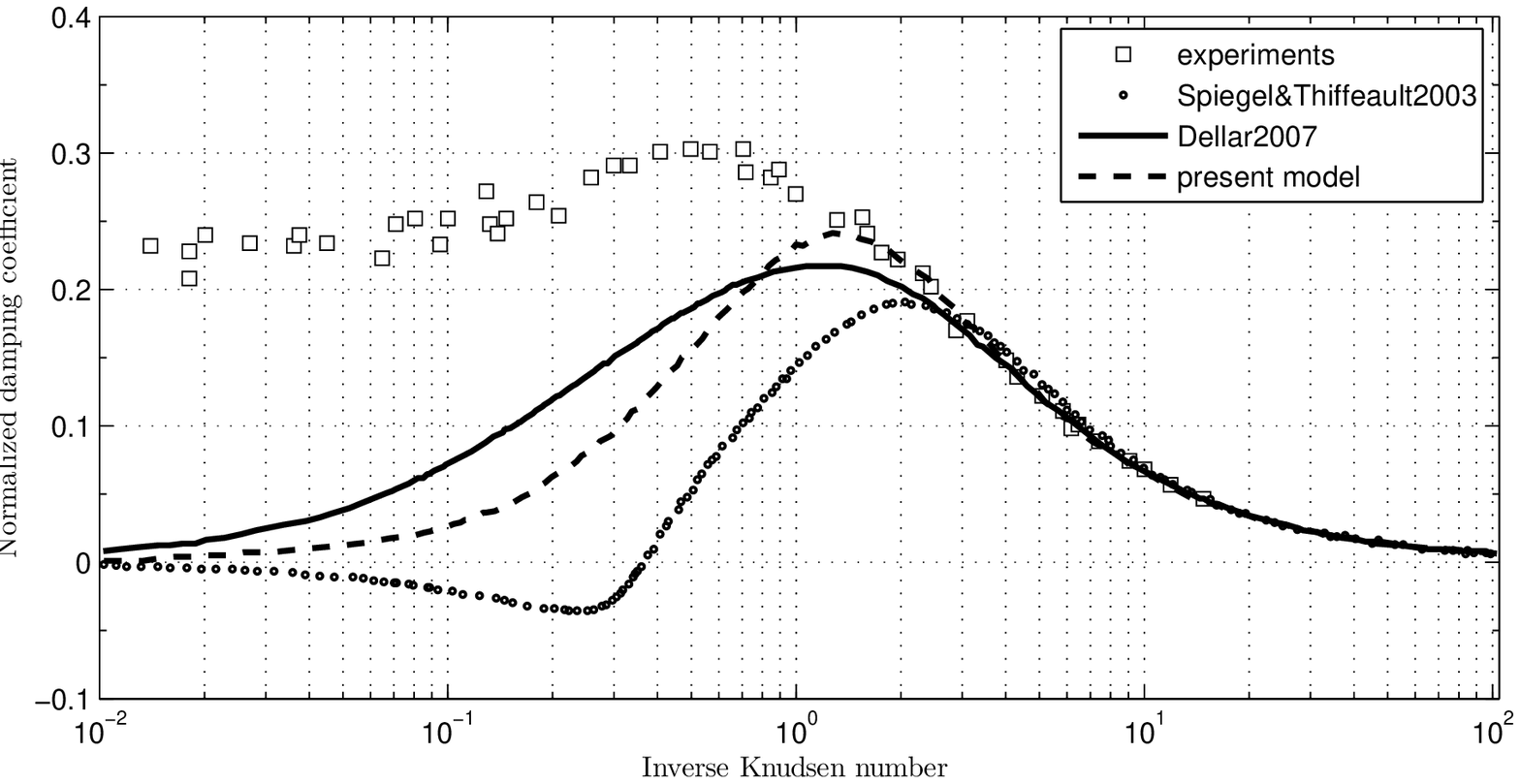}}\\
     \caption{Comparison of our volume based-model (as in figure \ref{kokou-second}) with two other recent models \cite{Spiegel_Thiffeault2003,Paul-Dellar.PhyofFlu2007}, and argon gas experimental data \cite{Meyer_Sessler1957}.}
     \label{comparedother}
\vspace{-10in}
\end{figure}

Now we set, $\alpha^* = \chi^*= 0$, that is, $W$ is given by an
expression with only the second order terms of equation
(\ref{expres_W}). In this case, we observe that the set of
volume-based equations has a wider range of stability, provided
$0\leq \gamma^* -\beta^*\leq 1.3$ approximately (see figure
\ref{stab-kokou-second}). Figure \ref{kokou-second} shows that the
phase speed prediction of one of the modes now agrees perfectly with
experiment, in both the low and the high frequency regimes. This
mode actually corresponds to the pressure mode, and it merges into
the new mass-density mode in the high frequency regime. For
comparison, in figure  \ref{comparedother} this physical mode is
plotted with the experimental data and results from two recent
continuum models derived as approximation solutions to the Boltzmann
equation \cite{Spiegel_Thiffeault2003,Paul-Dellar.PhyofFlu2007}. We
observe that our volume model is competitive with the best of these
two models. Our new model has the best damping coefficient
predictions in the low Knudsen number regime, and we note an
unphysical negative damping coefficient predicted by the second
order model of Spiegel and Thiffeault \cite{Spiegel_Thiffeault2003}.

In our investigations, our choice of the values of different
coefficients in the volume model has been primarily motivated by
finding the best agreement with the experimental data. However,
coefficient $S_c$, set to $0.9$ for figure \ref{navier-corrected},
agrees with an interpretation of $S_c$ as a Schmidt number with a
value of $5/6$ for monatomic hard sphere molecular gases; a value of
$0.75$ has been used for the dispersion analysis in reference
\cite{Marques_ChinePhyLet_2008}. While the stability depends on the
expression of $W$, our volume-based set of equations seems to remain
stable for whatever value the Schmidt number is set to, i.e.,
whatever the mass-density diffusivity.

The dimensionless expansion and compressibility coefficients we
obtained depart from their (equilibrium state) ideal gas values of
$1$. These departures from ideality may be attributable to real gas
effects now incorporated in our volume-based description. Similar
results to those presented in our figures are also obtained with
other combinations of the various coefficients. For example,
$\alpha^* =0.3$, $\chi^* =0.7$ and $S_c=3.33$ give the same results
as in figure \ref{navier-corrected}. This recalls experimental
reports that different gases can produce similar results
\cite{Meyer_Sessler1957,Greenspan_1950}. In any case, the various
coefficients in our volume model leave room to incorporate the
various properties of the gas under investigation.


\subsection{A prediction of the damping coefficient in the high frequency regime}
In figures \ref{damp-navier}, \ref{damp-navier-corrected} and
\ref{damp-kokou}, the predicted damping coefficient tends to zero as
the Knudsen number becomes large. This is a very common result when
using continuum models, as seen on figure \ref{comparedother}.
Problems have also been pointed out in comparisons with experiments
in this regime \cite{cercignani_rouge,cercignani-complet}.
Therefore, researchers have argued on the basis of spectral analysis
that continuum models based on a finite set of partial differential
equations cannot capture this branch of the graph
\cite{Paul-Dellar.PhyofFlu2007}. In any case, interpreting sound
waves in terms of pressure waves and momentum exchanges between
(only) molecules during collisions should be expected to lead to
vanishing damping as intermolecular collisions are no longer the
dominant phenomena in the very high Knudsen number regime
\cite{schotter:1974,Kanhn_Mintzer_1965_PhyofFlu}.

We now consider earlier comments by some investigators
\cite{schotter:1974,Maidanik_Fox_1965} who, analyzing the
experimental set-up, suggested that a model to predict this sound
dispersion must have a Knudsen number expression and a dimensional
analysis  that reflects the distinction between the
molecule/molecule collision-dominated regime and the
molecule/surface collision-dominated regime.

In the experimental set-up the gas was placed between source and
receiver then disturbed by a plane harmonic sound wave with a fixed
frequency at the source
\cite{Greenspan_1948,schotter:1974,Meyer_Sessler1957}. The primary
variable parameter in the experiments was the distance between the
source and the receiver. At very low pressures, the
molecule/molecule collisions that predominate in a high pressure (or
continuum) regime, become negligible, and molecular collisions with
surfaces dominate. In this situation, the microscopic collision
length scale becomes the distance traveled by molecules to reach the
surfaces --- no longer the mean free path that is the length scale
in the continuum regime. Accordingly, Schotter \cite{schotter:1974},
who also reported similar data to Greenspan, Meyer and Sessler,
presents a different dimensional analysis, introducing two different
microscopic times leading to two different Knudsen number
expressions. The first of these corresponds to a pressure-based
intermolecular collision time, and is the same definition as in
references \cite{Greenspan_1950,Meyer_Sessler1957}. The second
microscopic time is independent of molecule/molecule momentum
transfers and instead characterizes the frequency of collisions with
the surfaces. As we show explicitly in the Appendix, Greenspan's
dimensionless quantities in equation (\ref{damp-speed-expre}), and
the accompanying interpretation of frequency as a (conventional)
Knudsen number, are founded on molecule/molecule collisions and so
become inappropriate at high Knudsen number where these types of
collisions are no longer the principal momentum transfer mechanism
(see also reference \cite{Maidanik_Fox_1965}). A dimensional
analysis using the separation distance between the surfaces leads to
a different expression for the dimensional damping coefficient in a
low pressure gas, which is also, conversely, invalid for high
pressure cases (i.e.\ at low conventional Knudsen number). This
second expression may also be derived using the following
observation.

In section \ref{sec-Linearequation} we performed a dimensional
analysis, and introduced equation (\ref{harmonic-wrong}) which
assumes the harmonic wave form. As the set of partial differential
equations is linearized and dimensionless, characteristic time and
length scales have therefore been introduced before equation
(\ref{harmonic-wrong}). A better way of expressing the harmonic wave
is in a completely dimensionless form, i.e.,
\begin{equation}
\phi^* =\phi_a^*  \exp\left[i\left( \omega^* t^* - K^* x^* \right) \right],
\label{harmonic-correct}
\end{equation}
where $\omega^*$ and $K^*$ are, respectively, the dimensionless
complex wave frequency and dimensionless wave number. Moreover,
$\omega^*=\omega \tau$ and $K^* =LK$, with $\tau$ and $L$ the
characteristic time and length previously introduced in equation
(\ref{micro-timedefo}). The constant coefficient $\sqrt{5/3}$, from
the adiabatic exponent of a monatomic gas, could be simply
incorporated in the definition of the reference speed and is not
here the main issue. The dimensionless phase speed and dimensionless
spatial damping coefficient are therefore:
\begin{align}\label{damp-speed-expre-correct}
  \frac{1}{\Upsilon}_h  =  \sqrt{\frac{5}{3}}\frac{Re[K^*]}{\omega^*} ,  \ \  \Lambda_h = -  \sqrt{\frac{5}{3}}Im[K^*] \ ,
\end{align}
and we observe that while the dimensionless phase speed remains the
same as previously, the dimensionless damping coefficient is
different (see equation \ref{damp-speed-expre}): it does not contain
the frequency.


\begin{figure}[p]
     \begin{center}
           \includegraphics[width=0.8\textwidth,height=0.4\textwidth]{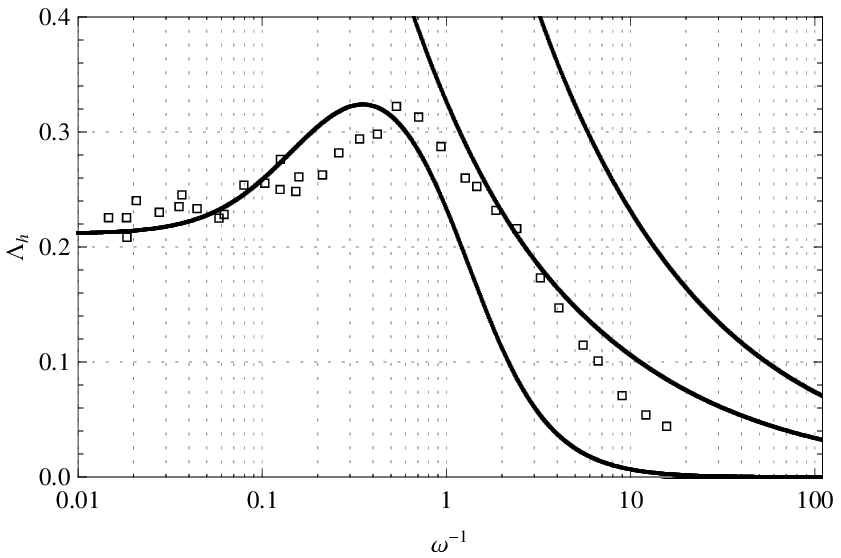}
     \caption{Damping coefficient predictions with $W$ described by a second
     order approximation, and using the definitions in equation
     (\ref{damp-speed-expre-correct}); $\beta^*=0.28$, $\gamma^*=0.48$, $S_c=0.14$.
     Note the agreement with one of the modes at high Knudsen numbers.}
     \label{paradoxical_damp}
\end{center}
\end{figure}

In figure \ref{paradoxical_damp} we plot the dimensionless damping
coefficient by our new hydrodynamic model, but using the redefined
expressions in equation (\ref{damp-speed-expre-correct}) (and using
same coefficients $S_c$, $\beta^*$ and $\gamma^*$ as in figure
\ref{kokou-second}). It is seen that our model reproduces the high
frequency branch, with the correct asymptotic value of the damping.
In addition, this is represented by the new mass-density mode, not
the classical pressure mode which instead diverges. Broadly, this
curve catches the shape and the shallow maximum around $Kn\approx
1$. The agreement is not so good by $Kn=1$, and becomes somewhat
inaccurate for low Knudsen numbers, as expected.

In summary, expressions (\ref{damp-speed-expre}) and
(\ref{damp-speed-expre-correct}) are each compatible with different
Knudsen number regimes and are both required for a proper
interpretation of the experimental results. Our volume-based
hydrodynamic model has been shown, therefore, to predict both the
low and the high frequency branch of the damping coefficient well,
while the inverse phase speed is always well-predicted.

\begin{figure}[p]
     \begin{center}
           \includegraphics[width=0.8\textwidth,height=0.4\textwidth]{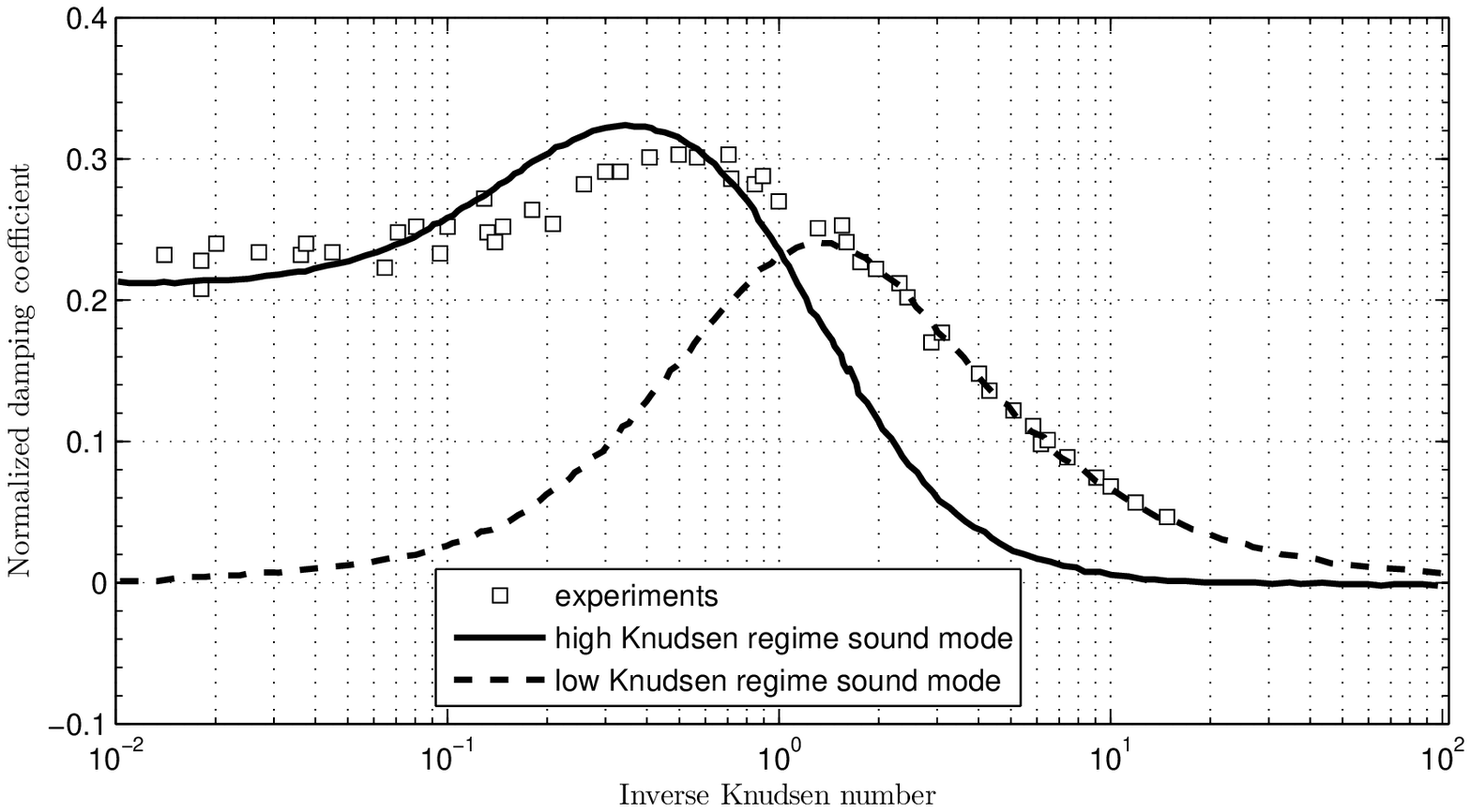}
     \caption{The two different natures of the sound mode, illustrated by the two
     different modes fitting experimental damping in different Knudsen number regimes.}
     \label{dampingkokouall}
\end{center}
\end{figure}

In his experiments, Schotter  \cite{schotter:1974} reported plane
standing waves for all Knudsen numbers. Because of difficulties
surrounding the predictions of the high Knudsen number branch, other
researchers assumed, however, that a plane wave analysis could not
capture this regime
\cite{Paul-Dellar.PhyofFlu2007,Kanhn_Mintzer_1965_PhyofFlu,cercignani_rouge}.
In our analysis, mass-density and pressure fields are plane harmonic
and therefore agree also with Schotter's experimental observation.
We also confirm the unusual (i.e.\ non-pressure-wave)
characteristics of sound waves in this regime because our good
predictions here are provided by our model's mass-density diffusion
terms. This is illustrated in figure \ref{dampingkokouall}, where
the two different modes fitting the experimental damping data in the
low and the high frequency regimes are both plotted.

Finally, even with the modified definitions of equation
(\ref{damp-speed-expre-correct}), the Navier-Stokes-Fourier model
gives at $1/Kn=0.01$ a value of the damping which is $30$ times the
experimental value of approximately $0.2$. So the conventional model
still provides incorrect predictions.

\section{Discussion}
Predicting sound wave phase speed and damping is a challenge both
for kinetic models derived from the Boltzmann dilute gas equation
and for continuum fluid hydrodynamics \cite{cercignani-complet}. The
few kinetic models
\cite{Buckner_PhyFlu1966,Sirovich_thurber1965,Kanhn_Mintzer_1965_PhyofFlu}
that agree with the experimental data over the entire range of
Knudsen number suffer three major criticisms. First, questions often
arise about the compatibility of kinetic boundary value problems
with experimental measurement
\cite{cercignani_rouge,cercignani-complet}. Second, the kinetic
models predict non-standard pressure fields
\cite{Buckner_PhyFlu1966}; in contrast, experiments have been based
on harmonic pressure waves, and indicate a plane standing wave
existing in the gas medium at all Knudsen numbers during
measurement. Third, the different mechanisms of momentum transfer in
the high pressure and the low pressure cases are not always
compatible with the kinetic model predictions
\cite{Maidanik_Fox_1965,Buckner_PhyFlu1966,cercignani-complet}. A
final issue, often raised with continuum fluid models beyond
Navier-Stokes-Fourier, is the appearance of a large number of modes
so it is not always easy to identify the mode that should describe
the sound wave.

Our figures \ref{kokou-second} and \ref{paradoxical_damp} show that
the continuum-based model considered in this paper reproduces the
experiments over the range of Knudsen number without the
difficulties listed above. In these figures there are only three
distinct modes to be associated with pressure, temperature and
mass-density in a given regime. In our understanding, pressure and
mass-density disturbances are distinct plane harmonic waves that
dominate in different Knudsen number regimes (see figure
\ref{dampingkokouall}). The existence of a mass-density wave
explains the plane standing wave observed in experiments in the high
Knudsen number regime; this mode is non-existent in conventional
fluid dynamic equations as there is no explicit mass-density
diffusion (or mass-density wave propagation). The agreement between
our theoretical damping results and experiment can be fully
explained in terms of two mean-free-paths inherent in the
experimental set-up; one mean free path is founded on the standard
kinetic pressure and molecular collisions, and the other founded on
the separation distances of the solid surfaces. The latter also
underlines the fundamental basis of our new approach itself: the
variation of the surface position is easily associated with
variation of the volume between molecules.

Our prediction of the high Knudsen number regime is possible only if
we adopt the second-order expression for $W$ given in equation
(\ref{expres_W}). This shows that this regime is best described by
microscopic structure evolutions, and not macroscopic average
thermodynamic property evolutions; therefore there is no localized
thermodynamic equilibrium in this case. Indeed, in equation
(\ref{approx-property-fun}) the time rate of change of the
microscopic volume $v$ is represented by the sum of the time rate of
change of the average value $\bar{v}$ and the change in its random
component, which is approximated using a relaxation time.
Consequently, the second-order terms involved in equation
(\ref{expres_W}) can be considered expressions of the random
component of the microscopic volume evolutions. (A representation of
microscopic structure, as in equation (\ref{approx-property-fun}),
is common in ``fading memory'' concepts, where it is given generally
as a convolution function
\cite{Ingo_muller_2001,Petrov_Szekeres_2008}.)

\section{Conclusion}
The starting point of our volume-based hydrodynamic model is the
representation of the fluid mass-density within conventional
continuum fluid mechanics and kinetic theory \cite{RGD-dadzie-2008}.
In this paper, we have seen that a volume-modified hydrodynamic
model can achieve surprisingly good results for sound wave
dispersion in monatomic gases. This problematic gas flow in the
non-continuum regime has previously been classified as
non-predictable using a continuum-based description. Moreover, our
volume-based hydrodynamics offers a more plausible interpretation of
the experimental data than some previous kinetic results.

We therefore propose the volume-based model for further
investigations. First, more sophisticated constructions of the new
volume variation terms involved in the description are required, as
results suggest some sensitivities to their formulation. Second,
further application should be made to other flows and heat transfer
configurations where the classical continuum models become
inadequate. For example, investigating heat transfer in the
transition regime, where the dependency of heat conductivity on the
Knudsen number or pressure, and the definition of heat flux, are
still unresolved problems \cite{mandell-1924}.

\section*{Acknowledgements}
This work is funded in the UK by the Engineering and Physical
Sciences Research Council under grant EP/D007488/1, and through a
Royal Society of Edinburgh / Scottish Government Support Research
Fellowship for JMR. The authors would like to thank the referees of
this paper for their helpful comments.

\bibliographystyle{elsart-num}

\begin{thebibliography}{10}
\expandafter\ifx\csname url\endcsname\relax
  \def\url#1{\texttt{#1}}\fi
\expandafter\ifx\csname urlprefix\endcsname\relax\def\urlprefix{URL
}\fi

\bibitem{DadzieReese.PhysicaA.2008}
S.~K. Dadzie, J.~M. Reese, C.~R. McInnes, A continuum model of gas
flows with
  localized density variations, Physica A 387~(24) (2008) 6079--6094.

\bibitem{LockerbyReeseStruchtrup.ProcRoySoc.2009}
D.~A. Lockerby, J.~M. Reese, H.~Struchtrup, Switching criteria for
hybrid
  rarefied gas flow solvers, Proceedings of the Royal Society A 465 (2009)
  1581--1598.

\bibitem{cercignani_rouge}
C.~Cercignani, Rarefied Gas Dynamics: from {B}asic {C}oncepts to
{A}ctual
  {C}alculations, Cambridge University Press, 2000.

\bibitem{cercignani-complet}
C.~Cercignani, Theory and Application of the {B}oltzman Equation,
Scottish
  Academic Press, 1975.

\bibitem{Greenspan_1950}
M.~Greenspan, Propagation of sound in rarefied helium, Journal of
the
  Acoustical Society of America 22~(5) (1950) 568--571.

\bibitem{Meyer_Sessler1957}
E.~Meyer, G.~Sessler, Schallausbreitung in gasen bei hohen
frequenzen und sehr
  niedrigen drucken, Zeitschrift fur Physik 149 (1957) 15--39.

\bibitem{Kanhn_Mintzer_1965_PhyofFlu}
D.~Kahn, D.~Mintzer, Kinetic theory of sound propagation in rarefied
gases,
  Physics of Fluids 8~(6) (1965) 1090--1102.

\bibitem{StruchtrupTorrilhon2003PhyFlu}
H.~Struchtrup, M.~Torrilhon, Regularization of {G}rad's 13 moment
equations:
  derivation and linear analysis, Physics of Fluids 15~(9) (2003) 2668.

\bibitem{schotter:1974}
R.~Schotter, Rarefied gas acoustics in the noble gases, Physics of
Fluids
  17~(6) (1974) 1163--1168.

\bibitem{Buckner_PhyFlu1966}
J.~K. Buckner, J.~H. Ferziger, Linearized boundary value problem for
a gas and
  sound propagation, Physics of Fluids 9~(12) (1966) 2315--2322.

\bibitem{Paul-Dellar.PhyofFlu2007}
P.~J. Dellar, Macroscopic descriptions of rarefied gases from the
elimination
  of fast variables, Physics of Fluids 19~(10) (2007) 107101.

\bibitem{Sirovich_thurber1965}
L.~Sirovich, J.~K. Thurber, Propagation of forced sound waves in
rarefied
  gasdynamics, Journal of the Acoustical Society of America 37~(2) (1965)
  329--339.

\bibitem{Loyalka_Cheng_PhysFlu_1979}
S.~K. Loyalka, T.~C. Cheng, Sound-wave propagation in rarefied gas,
Physics of
  Fluids 22~(5) (1979) 830--836.

\bibitem{Spiegel_Thiffeault2003}
E.~A. Spiegel, J.-L. Thiffeault, Higher-order continuum
approximation for
  rarefied gases, Physics of Fluids 15~(11) (2003) 3558--3567.

\bibitem{Cattaneo_CRAS_1958}
C.~Cattaneo, A form of heat conduction equation which eliminates the
paradox of
  instantaneous propagation, Comptes Rendus de l'Academie des Sciences 247
  (1958) 431--433.

\bibitem{Ingo_muller_2001}
I.~Muller, Extended thermodynamics -- the physics and mathematics of
the
  hyperbolic equations of thermodynamics, International Series of Numerical
  Mathematics 141 (2001) 733.

\bibitem{Petrov_Szekeres_2008}
N.~Petrov, A.~Szekeres, New approach to the non-classical heat
conduction,
  Journal of Theoretical and Applied Mechanics, Sofia 38~(3) (2008) 61--70.

\bibitem{zhuomin2007}
Z.~M. Zhang, Nano Microscale Heat Transfer, McGraw-Hill
Professional, 2007.

\bibitem{Brenner-Phys.vol2005}
H.~Brenner, Kinematics of volume transport, Physica A 349~(1-2)
(2005) 11--59.

\bibitem{Brenner.PhysicaA.revs.2005}
H.~Brenner, Navier-{S}tokes revisited, Physica A 349~(1-2) (2005)
60--132.

\bibitem{Marques_ChinePhyLet_2008}
W.~M. Jr, Is {B}renner's modification to the classical
{N}avier-{S}tokes
  equations able to describe sound propagation in gases?, Chinese Physics
  Letters 25~(4) (2008) 1355--1358.

\bibitem{Greensield_reese_2007}
C.~J. Greenshields, J.~M. Reese, The structure of shock waves as a
test of
  {B}renner's modifications to the {N}avier-{S}tokes equations, Journal of
  Fluid Mechanics 580 (2007) 407--429.

\bibitem{Mario_PRL}
M.~Liu, Comments on ``{W}eakly and strongly consistent formulations
of
  irreversible processes'', Physical Review Letters 100~(9) (2008) 098901.

\bibitem{Maidanik_Fox_1965}
G.~Maidanik, H.~L. Fox, Comments on ``{P}ropagation of forced sound
waves in
  rarefied gasdynamics'', Journal of the Acoustical Society of America 38~(3)
  (1965) 477--478.

\bibitem{Greenspan_1948}
M.~Greenspan, Attenuation of sound in rarefied helium, Physical
Review 75~(1)
  (1948) 197--198.

\bibitem{RGD-dadzie-2008}
S.~K. Dadzie, J.~M. Reese, The concept of mass-density in classical
  thermodynamics and the {B}oltzmann kinetic equation for dilute gases, in:
  Rarefied Gas Dynamics: 23rd International Symposium, Vol. 1084, 2008, p. 117.

\bibitem{mandell-1924}
W.~Mandell, J.~West, On the temperature gradient in gases at various
pressures,
  {P}roceedings of the {P}hysical Society of {L}ondon 37~(1) (1924) 20--41.

\end{thebibliography}

\appendix

\subsection*{Analysis of Greenspan's interpretation of Knudsen number
variations}
This is a boundary value problem, with $w$  positive
real, and $K=(K_r+iK_i)$ a complex number. A plane harmonic wave
$\phi (t,x)$ is written with dimensional variables as
\begin{equation}
\phi (t,x) = \exp\left[i\left( \omega t - (K_r+iK_i) x \right) \right].
\label{app-harmonic1}
\end{equation}
We seek dimensionless expressions for the phase speed and damping.
First, equation (\ref{app-harmonic1}) is rewritten,
\begin{equation}
\phi(t,x) = \exp\left[i \omega \left( t - \frac{K_r}{\omega} x \right) \right] \exp\left[\left(\frac{K_i}{\omega}\right) \omega x\right].
\label{app-harmonic2}
\end{equation}
The experimental set-up infers a fixed frequency, $w_e$
\cite{Meyer_Sessler1957,Greenspan_1950,schotter:1974}. Suppose that
the gas has well-defined microscopic time and length scales, $\tau$
and $L$, respectively, which therefore specify a microscopic  speed
$C_0$. We may then define dimensionless frequency, time and length
as
\begin{align}\label{app-micro-timedefo}
\omega = \omega_e \omega^*, \  \  t =\tau t^*= \frac{L}{C_0} t^*, \ \  x = L x^* .
\end{align}
Using these definitions, equation (\ref{app-harmonic2}) becomes,
\begin{equation}
\phi(t,x) = \exp\left[i \omega^* \omega_e \tau\left( t^* - \frac{K_r}{\omega} C_0 x^* \right) \right] \exp\left[C_0\frac{K_i}{\omega} \omega^*\omega_e \tau x^*\right].
\label{app-harmonic3}
\end{equation}
Away from any gas/surface interaction, the mean free time describing
the average collision time between two molecules is well-defined. We
may therefore choose $\tau$ to be the time between successive
molecular collisions. In such a case, and with $\omega_e$ defining
the flow macroscopic time scale, we have a Knudsen number
$Kn=\omega_e\tau$. Subsequently, equation (\ref{app-harmonic3})
yields,
\begin{equation}
\phi(t,x) = \exp\left[i \omega^* Kn \left( t^* - \frac{K_r}{\omega} C_0 x^* \right) \right] \exp\left[C_0\frac{K_i}{\omega} \omega^*Kn x^*\right].
\label{app-harmonic4}
\end{equation}
We therefore have a dimensionless inverse speed $C_0 K_r/\omega$ and
a dimensionless damping coefficient $-C_0K_i/ \omega$. Meanwhile,
the dimensionless frequency is a product: $\omega^*Kn$. This means
that for a fixed value of $Kn$, the Knudsen number is a simple
scaling factor for the dimensionless frequency. Conversely, a fixed
value of the dimensionless frequency is a simple scaling factor for
the Knudsen number. Consequently, and for this particular
configuration, one may absorb the factor $Kn$ into $\omega^*$ and
interpret the variation of their product as either Knudsen number or
dimensionless frequency variations.

However, this description relies on the definition of the
microscopic time $\tau$ as the time between molecule/molecule
collisions. If this microscopic time is physically undefined, or
becomes large, then equation (\ref{app-harmonic4}) and the
interpretation that follows it becomes invalid because the product
$\omega^*Kn$ is indeterminate. This is the case when the gas is
confined between two surfaces so that collisions between molecules
are no longer the most important mechanism of momentum transfer from
one surface to the other, and instead the interactions of the
molecules directly with the two surfaces (the source and receiver in
the experiments) is.

In Greenspan's work, which has been followed by several authors, the
non-dimensionalisation starts with a reference speed, denoted
$v_0=w/\beta_0$, which in our notation corresponds to $w/C_0$,
assuming an approximation of the dispersion at high pressure. Then
the intermolecular collision mean time $\tau$ is determined assuming
Maxwell molecules. The dimensionless sound speed and damping are
given as they appear through equation (\ref{app-harmonic4}) while
the inverse of the product $\omega^*Kn$ is referred to as
``Reynold's number'' .

In any case, one can see easily from the expression $C_0K_i/ \omega$
that for all theories predicting a finite value of the damping this
dimensionless expression should give zero damping for $\omega$
tending to infinity. So, the expression, at first glance, is not
even a well-indicated form to compare between different theoretical
results in this field. A different analysis is therefore required.

Returning to equation (\ref{app-harmonic1}), for high Knudsen
numbers let us assume that the separation distance between the two
surfaces, $L$, is the relevant microscopic parameter. With a $C_0$
that may be the thermal speed (or any other characteristic molecular
speed), the average time spent travelling between the surfaces is
now associated with $\tau$ \cite{schotter:1974}. As there are, on
average,  no intermolecular collisions in that period we expect the
wave propagation to become independent of the conventional Knudsen
number beyond a certain limit. Equation (\ref{app-harmonic1}) is
then written,
\begin{equation}
\phi(t,x) = \exp\left[i\left( \omega \tau t^* - L(K_r+iK_i) x^* \right) \right],
\label{app-harmonic12}
\end{equation}
which implies $\omega^*=\omega \tau$, $K^*= LK$, and the
dimensionless sound speed and damping are given, respectively, by
$\omega^*/K_r^* $ and $-K_i^*$, which are the expressions we defined
in equation (\ref{harmonic-correct}) (allowing for the constant
coefficient $\sqrt{5/3}$). Moreover, this dimensionless phase speed
and damping are independent of the dimensional frequency $\omega$
and  so independent of $\omega_e$.

Although our corrected dimensional analysis seems to work with the
data in reference \cite{Meyer_Sessler1957}, further verifications
with other experiments using reliable dimensionless parameters are
necessary. It is also worth noting that the failure of Greenspan's
analysis at high frequencies means that a high conventional Knudsen
number does not necessarily mean a high frequency, and \textit{vice
versa}. In Figure \ref{paradoxical_damp}, $\omega$ is strictly
speaking referring to a separation-distance-based Knudsen number,
not the real dimensional frequency --- as we have shown through
equation (\ref{app-harmonic12}).

We have not compared our theoretical results with the more recent
experimental data by Schotter \cite{schotter:1974}. This is because,
while Schotter differentiated between two microscopic time scales,
he defined the dimensionless parameters as in Greenspan's analysis,
i.e., a dimensionless damping coefficient that depends on the
frequency over the full regime. He reported different plots for
different separation distances.

\newpage

\end{document}